\documentclass{wqcd03}                 

\usepackage{txfonts}                   
\confname{QCD@Work 2003 - International Workshop on QCD, Conversano, Italy, 
14--18 June 2003}

\title{Loops and Power Counting in the High Density Effective Field Theory}

\author{Thomas Sch{\"a}fer\addressmark{a,b}}

\address[a]{Department of Physics, North Carolina State University,
Raleigh, NC 27695}
\address[b]{Riken-BNL Research Center, Brookhaven National Laboratory,
Upton, NY 11973}
\newcommand {\bea}{\begin{eqnarray}}
\newcommand {\eea}{\end{eqnarray}}
\newcommand {\be}{\begin{equation}}
\newcommand {\ee}{\end{equation}}

\begin{document}

\begin{abstract}
We discuss the high density effective theory of QCD. We concentrate
on the problem of developing a consistent power counting scheme.
\end{abstract}

\maketitle


\section{Introduction}
\label{sec_intro}

  The study of hadronic matter in the regime of high baryon density 
has led to the theoretical prediction of several new phases of strongly 
interacting matter, such as color superconducting quark matter and 
color-flavor locked matter \cite{Bailin:1984bm,Alford:1998zt,Rapp:1998zu,Alford:1999mk,Schafer:1999fe,Rajagopal:2000wf,Alford:2001dt,Schafer:2003vz,Rischke:2003mt}. 
These phases may be realized in nature in the cores of neutron
stars. In order to study this possibility quantitatively we would 
like to develop a systematic framework that will allow us to determine 
the exact nature of the phase diagram as a function of the density, 
temperature, the quark masses, and the lepton chemical potentials, 
and to compute the low energy properties of these phases. 

  If the density is large then the Fermi momentum is much bigger 
than the QCD scale, $p_F=\mu\gg\Lambda_{QCD}$, and asymptotic
freedom implies that the effective coupling is weak. It would
then seem that such a framework is provided by weak perturbative 
QCD. It is well known, however, that a naive expansion in powers 
of $\alpha_s$ is not sufficient. Long range gauge boson exchanges 
lead to infrared divergencies that require resummation. In a 
degenerate Fermi system the effect of the BCS or other pairing 
instabilities have to be taken into account. And finally, in 
systems with broken global symmetries, the low energy properties 
of the system are governed by collective modes that carry the 
quantum numbers of the broken generators. 

  In order to address these problems it is natural to exploit 
the separation of scales provided by $\mu\gg g\mu \gg\Lambda_{QCD}$ 
in the normal phase, or $\mu\gg g\mu \gg\Delta\gg\Lambda_{QCD}$ in 
the superfluid phase. An effective field theory approach to 
phenomena near the Fermi surface was suggested by Hong
\cite{Hong:2000tn,Hong:2000ru}. This approach was applied
to a number of problems \cite{Beane:2000ji,Beane:2000ms,Schafer:2001za,Casalbuoni:2001ha}, see \cite{Nardulli:2002ma} for a review.
Even though a number of interesting results have been obtained
there are a number of important conceptual issues that are not 
very well understood. These issues concern power counting,
renormalization and matching. In this contribution we would like to 
study some of these issues in more detail. 

\section{High Density Effective Theory (HDET)}
\label{sec_hdet}

\begin{figure}
\hbox to\hsize{\hss
\includegraphics[width=6cm]{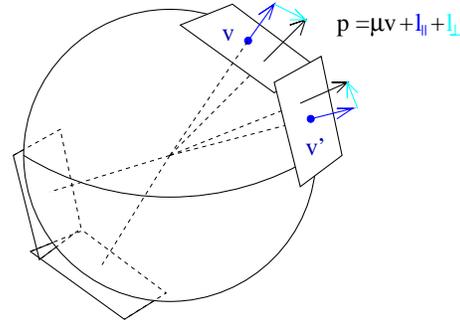}
\hss}
\caption{High density effective field theory description of 
excitations near the Fermi surface. The effective 
theory is defined on patches labeled by the local
Fermi velocity $v$. Momenta are decomposed with 
respect to $v$, $\vec{p}=\mu\vec{v}+l_\perp+l_{||}$.}
\label{fig_fs}
\end{figure}

 At high baryon density the relevant degrees of freedom are 
particle and hole excitations which move with the Fermi 
velocity $v$. Since the momentum $p\sim v\mu$ is large, 
typical soft scatterings cannot change the momentum by very 
much. An effective field theory of particles and holes 
in QCD is given by \cite{Hong:2000tn,Hong:2000ru,Beane:2000ms}
\be
\label{l_hdet}
{\cal L} =\sum_{v}
 \psi_{v}^\dagger (iv\cdot D) \psi_{v} 
 -\frac{1}{4}G^a_{\mu\nu} G^a_{\mu\nu}+ \ldots ,
\ee
where $v_\mu=(1,\vec{v})$. The field describes particles and 
holes with momenta $p=\mu\vec{v}+l$, where $l\ll\mu$. We will 
write $l=l_0+l_{\|}+l_\perp$ with $\vec{l}_{\|}
=\vec{v}(\vec{l}\cdot \vec{v})$ and $\vec{l}_\perp = 
\vec{l}-\vec{l}_{\|}$. In order to take into account the entire 
Fermi surface we have to cover the Fermi surface with patches 
labeled by the local Fermi velocity, see Fig.~\ref{fig_fs}. The 
number of such patches is $n_v\sim (\mu^2/\Lambda_\perp^2)$ where 
$\Lambda_\perp \ll\mu$ is the cutoff on the transverse momenta 
$l_\perp$. 

 Higher order terms are suppressed by powers of $1/\mu$. As 
usual we have to consider all possible terms allowed by the 
symmetries of the underlying theory. At $O(1/\mu)$ we have
\be
{\cal L} = \sum_{v}\left\{
  -\frac{1}{2\mu} \psi_{v}^\dagger D_\perp^2 \psi_{v}
  - ag  \psi_{v}^\dagger \frac{\sigma^{\mu\nu} G_{\mu\nu}^\perp}
   {4\mu}\psi_v \right\}. 
\ee
The coefficient of the first term is fixed by the dispersion 
relation of a fermion near the Fermi surface, $l_0=l_{\|}+
l_\perp^2/(2\mu)+\ldots$. The coefficient of the second term is 
most easily determined by integrating out anti-particles at 
tree level. We find $a=1+O(g^2)$, where the $O(g^2)$ terms 
arise from higher order perturbative corrections. At higher 
order in $1/\mu$ there is an infinite tower of operators 
of the form $\mu^{-n}\psi^\dagger_v D_\perp^{2n_1}(\bar{v}
\cdot D)^{n_2} \psi_v$ with $\bar{v}=(1,-\vec{v})$ and $n=
2n_1+n_2-1$. 

\begin{figure}
\hbox to\hsize{\hss
\includegraphics[width=\hsize]{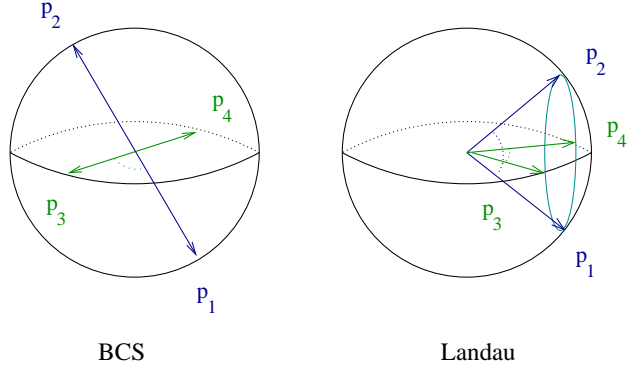}
\hss}
\caption{Kinematics of four-fermion operators in the
effective theory.}
\label{fig_4f_kin}
\end{figure}

At $O(1/\mu^2)$ the effective theory contains four-fermion 
operators 
\bea 
{\cal L} &=&  \frac{1}{\mu^2} \sum_{v_i} \sum_{\Gamma,\Gamma'} 
  c^{\Gamma\Gamma'}(\vec{v}_1\cdot\vec{v}_2,\vec{v}_1\cdot\vec{v}_3,
   \vec{v}_2\cdot\vec{v}_3) \nonumber \\
  & & \mbox{}\hspace{0.5cm}\cdot 
   \Big(\psi_{v_1} \Gamma \psi_{v_2}\Big)
   \Big(\psi^\dagger_{v_3}\Gamma'\psi^\dagger_{v_4}\Big)
   \delta(v_1+v_2-v_3-v_4).
\eea
The restriction $v_1+v_2=v_3+v_4$ allows two types of four-fermion 
operators, see Fig.~\ref{fig_4f_kin}. The first possibility is that 
both the incoming and outgoing fermion momenta are back-to-back. This 
corresponds to the BCS interaction
\be
\label{c_bcs}
{\cal L}=  \frac{1}{\mu^2}\sum_{v,v'}\sum_{\Gamma,\Gamma'}
  V_l^{\Gamma\Gamma'} R_l^{\Gamma\Gamma'}(\vec{v}\cdot\vec{v}') 
    \Big(\psi_{v} \Gamma \psi_{-v}\Big)
   \Big(\psi^\dagger_{v'}\Gamma'\psi^\dagger_{-v'}\Big),
\ee
where $\vec{v}\cdot\vec{v}'=\cos\theta$ is the scattering 
angle and $R_l^{\Gamma\Gamma'}(\vec{v}\cdot\vec{v}')$ is a 
set of orthogonal polynomials that we will specify below.
The second possibility is that the final momenta are equal 
to the initial momenta up to a rotation around the axis 
defined by the sum of the incoming momenta. The relevant 
four-fermion operator is 
\be
\label{c_flp}
{\cal L}=  \frac{1}{\mu^2}\sum_{v,v',\phi}\sum_{\Gamma,\Gamma'}
  F_l^{\Gamma\Gamma'}(\phi) R_l^{\Gamma\Gamma'}(\vec{v}\cdot\vec{v}') 
    \Big(\psi_{v} \Gamma \psi_{v'}\Big)
   \Big(\psi^\dagger_{\tilde{v}}\Gamma'\psi^\dagger_{\tilde{v}'}
  \Big),
\ee
where $\tilde{v},\tilde{v}'$ are the vectors obtained from 
$v,v'$ by a rotation around $v_{tot}=v+v'$ by the angle $\phi$.
In a system with short range interactions only the quantities 
$F_l(0)$ are known as Fermi liquid parameters. In QCD forward
scattering is correctly reproduced by the leading order HDET 
lagrangian, but exchange terms have to be absorbed into four 
fermion operators \cite{Hands:2003rw}.

  The matrices $\Gamma,\Gamma'$ describe the spin, color and flavor 
structure of the interaction. The spin structure is most easily discussed 
in terms of helicity amplitudes. As an example, we consider the BCS 
operators $(v,-v)\to(v',-v')$. The spins of the two quarks
can be coupled to total spin zero or one. In the spin zero 
sector there are two possible helicity channels, $(++)\to(++)$ 
and $(++)\to (--)$ together with their parity partners $(+
\leftrightarrow -)$. In the limit $m\to 0$ perturbative 
interactions only contribute to the helicity non-flip 
amplitude
\bea
\label{v_bcs_0}
{\cal L} &=&  \frac{1}{\mu^2}\sum_{v,v'}
  V_l^{++} P_l(\vec{v}\cdot\vec{v}') 
    \Big(\psi_{v} \sigma_2 H_+\psi_{-v}\Big)
   \Big(\psi^\dagger_{v'}\sigma_2 H_+\psi^\dagger_{-v'}\Big)
 \nonumber \\
& & \mbox{}\hspace{3cm}
 + (+\leftrightarrow -),
\eea
where $P_l(\cos\theta)$ are Legendre polynomials and $H_\pm$ are
helicity projectors. Quark mass terms as well as non-perturbative 
effects associated with instantons induce helicity flip operators 
\cite{Schafer:2001za,Schafer:2002ty}. These operators are suppressed 
by additional powers of $1/\mu$, but they have important physical 
effects. For example, helicity flip amplitudes determine the masses 
of Goldstone bosons in the CFL and 2SC phases. 

 In the spin one sector there is only one helicity channel 
$(+-)\to(+-)$. The corresponding BCS interaction is 
\bea
\label{v_bcs_1}
{\cal L} &=&  \frac{1}{\mu^2}\sum_{v,v'}
  V_l^{+-} d^{(l)}_{11}(\vec{v}\cdot\vec{v}') 
    \Big(\psi_{v} \sigma_2H_-\vec{\sigma} H_+\psi_{-v}\Big)
 \nonumber \\ 
 & & \mbox{}\hspace{1cm}\cdot
   \Big(\psi^\dagger_{v'}\sigma_2H_-\vec{\sigma} H_+
    \psi^\dagger_{-v'}\Big)
 + (+\leftrightarrow -),
\eea
where $d^{(l)}_{11}(\cos\theta)$ is the reduced Wigner 
D-function. In addition to the operators considered
here there is, of course, an infinite tower of operators 
with more fermion fields or extra covariant derivatives.

\section{Matching}
\label{sec_match}

\begin{figure}
\hbox to\hsize{\hss
\includegraphics[width=\hsize]{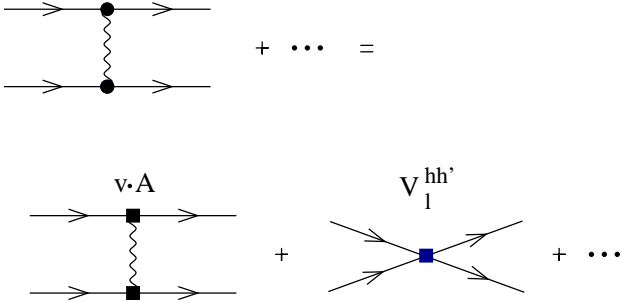}
\hss}
\caption{Matching condition for the four-fermion operators
in the high density effective theory.}
\label{fig_match}
\end{figure}

 The four-fermion operators in the effective theory can 
be determined by matching the quark-quark scattering amplitudes 
in the BCS and forward scattering kinematics. As an illustration 
we consider the leading order $O(g^2)$ BCS amplitude in the spin 
zero color anti-triplet channel. The matching condition is
\be
 \int d\theta f^{++}_{HDET}(\theta)P_l(\theta)  
 = \int d\theta f^{++}_{QCD}(\theta)P_l(\theta),
\ee
where $f^{++}(\theta)$ is the on-shell $(v,-v)\to(v',-v')$
scattering amplitude in the helicity $(++)\to(++)$ channel
as a function of the scattering angle $cos\theta=v\cdot v'$, 
see Fig.~\ref{fig_match}. The scattering amplitude in the
effective theory contains almost collinear gluon exchanges 
which do not change the velocity label of the quarks as well 
as four-fermion operators which correspond to scattering 
involving different patches on the Fermi surface. The 
collinear contribution to the moments of the scattering 
amplitude depends on the cutoff $\Lambda_\perp$ which 
we impose on the transverse momenta inside a given 
velocity patch. Since the moments in the microscopic
theory are independent of $\Lambda_\perp$ this dependence
has to cancel against the cutoff dependence of the 
coefficients of the four-fermion operators. 

 The matching condition is simplest for $\Lambda_\perp^2=
2\mu^2$. The s-wave term is given by $V^{++}_0(\Lambda_\perp^2
=2\mu^2) = 0$ up to corrections of $O(g^4)$ \cite{Schafer:2003jn}. 
The cutoff dependence of $V_0^{++}$ is controlled by the 
renormalization group equation 
\be 
\label{v0_rge}
\Lambda_\perp^2\frac{d}{d\Lambda_\perp^2}V_0^{++}(\Lambda_\perp^2)
 = \frac{g^2}{3}.
\ee
We can also compute the coefficients of four-fermion operators
corresponding to higher partial wave and operators with non-zero 
spin. For example, the angular momentum $l=1$ terms in the 
helicity zero and one channels are given by 
\bea 
V^{++}_1(\Lambda_\perp^2 = 2\mu^2) &=&
  -6\frac{g^2}{3} , \\
V^{+-}_1(\Lambda_\perp^2= 2\mu^2) &=&
  -\frac{9}{2}\frac{2g^2}{3} .
\eea
We will see below that these results determine the relative 
magnitude of the BCS gap in channels with different spin and 
angular momentum \cite{Brown:1999yd,Schafer:2000tw,Schmitt:2002sc}.

\section{Symmetries and Power Counting}
\label{sec_pow}

 In this section we shall discuss the symmetries of the high 
density effective theory and try to develop a systematic 
power counting. The high density effective theory has a
number of similarities with the heavy quark (HQET) and 
soft collinear (SCET) effective field theories. As in 
both of these theories, the fermion field is characterized 
by a velocity label, and the kinetic term is of the form
$v\cdot D$. Like SCET, HDET is a theory of ultra-relativistic 
particles and $v^2=0$. On the other hand, HDET has a
number of symmetries that are more akin to non-relativistic 
field theories. For example, the leading order HDET effective
lagrangian, equ.~(\ref{l_hdet}), has a SU(2) spin 
symmetry 
\be 
\psi_v \to \exp(i\vec{\theta}\cdot\vec{\sigma})\psi_v .
\ee
Superficially, terms that break this symmetry are 
suppressed by powers of $1/\mu$. We shall see that 
this is not true, however. Hard dense loops modify 
the power counting in HDET, and SU(2) violating terms
are not suppressed by powers of $\mu$, but only by powers 
of the coupling constant, $g$. The approximate spin symmetry 
of the HDET lagrangian has nevertheless important 
physical consequences. For example, to leading logarithmic
accuracy the BCS gap in the spin zero and spin one 
channel are the same. 

 The HDET lagrangian also possesses a reparametrization
invariance
\bea 
 \vec{v}&\to& \vec{v}+\vec{\epsilon}/\mu ,\\
 \vec{l}&\to& \vec{l}-\vec{\epsilon},\\
 \psi_v &\to& \psi_v+\delta\psi_v,
\eea
which reflects our freedom in choosing the local
Fermi velocity. Note that in order to keep $v^2=0$
we have to choose $\vec{v}\cdot\vec{\epsilon}=0$.
As usual, reparametrization invariance fixes the 
coefficients of certain higher order terms in the
effective lagrangian. 

 We now come to the issue of power counting. The 
power counting in HDET has a number of similarities
with the power counting in NRQCD and SCET, see 
\cite{Luke:1999kz,Bauer:2002uv} for a discussion of these
effective theories. We first discuss a ``naive''
attempt to count powers of the small scale $l$. In 
the naive power counting we assume that $v\cdot D$ scales
as $l$, $\psi_v$ scales as $l^{3/2}$, $A_\mu$ scales 
as $l$, and every loop integral scales as $l^4$. We 
also assume that $\vec{D}_\perp,\bar{v}\cdot D\sim l$. 
In this case it is easy to see that a general diagram 
with $V_k$ vertices of scaling dimension $k$ scales as
$l^\delta$ with
\be
\label{pc_naive}
\delta = 4 +\sum_k V_k(k-4).
\ee
A general vertex is of the form
\be
\label{dim_count}
\psi^a(v\cdot D)^b(\bar{v}\cdot D)^c (D_\perp)^d
 (1/\mu)^e,
\ee
and has mass dimension $3a/2+b+c+d-e=4$. Since $k=3a/2+b+c+d$
and $e\geq 0$ we have $k-4\geq 0$. This implies
that the power counting is trivial: All diagrams constructed
from the leading order lagrangian have the same scaling, 
all diagrams with higher order vertices are suppressed,
and the degree of suppression is simply determined by 
the number and the scaling dimension of the vertices.

\begin{figure}
\hbox to\hsize{\hss
\includegraphics[width=4cm]{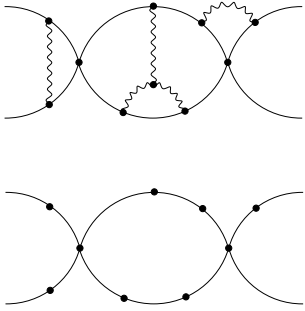}
\hss}
\caption{Counting hard loops in the effective field theory.
If all (soft) gluon lines are removed the remaining fermionic
loops contain sums over the velocity index.}
\label{fig_lcount}
\end{figure}

 Complication arise because not all loop diagrams
scale as $l^4$. In fermion loops sums over patches 
and integrals over transverse momenta can combine 
to give integrals that are proportional to the 
surface area of the Fermi sphere, 
\be
\label{hard_int}
 \frac{1}{2\pi}\sum_{\vec{v}}\int\frac{d^2l_\perp}{(2\pi)^2}
 =\frac{\mu^2}{2\pi^2}\int \frac{d\Omega}{4\pi}.
\ee
These loop integrals scale as $l^2$, not $l^4$. In the 
following we will refer to loops that scale as $l^2$
as ``hard loops'' and loops that scale as $l^4$ as ``soft 
loops''. In order to take this distinction into account we 
define $V_k^S$ and $V_k^H$ to be the number of soft and hard 
vertices of scaling dimension $k$. A vertex is called soft if it 
contains no fermion lines. In order to determine the $l$ 
counting of a general diagram in the effective theory we 
remove all gluon lines from the graph, see Fig.~\ref{fig_lcount}. 
We denote the number of connected pieces of the remaining graph 
by $N_C$. Using Euler identities for both the initial 
and the reduced graph we find that the diagram scales
as $l^\delta$ with 
\be
\label{pc_imp}
 \delta = \sum_i \left[ (k-4)V_k^S + (k-2-f_k)V_k^H\right]
 +E_Q +4 - 2N_C.
\ee
Here, $f_k$ denotes the number of fermion fields in  
a hard vertex, and $E_Q$ is the number of external quark 
lines. We observe that in general the scaling dimension 
$\delta$ still increases with the number of higher order 
vertices, but now there are two important exceptions. 

 First we observe that the number of disconnected fermion
loops, $N_C$, reduces the power $\delta$. Each disconnected
loop contains at least one power of the coupling 
constant, $g$, for every soft vertex. As a result, 
fermion loop insertions in gluon $n$-point functions 
spoil the power counting if the gluon momenta satisfy
$l\sim g\mu$. This implies that for $l<g\mu$ the high 
density effective theory becomes non-perturbative and
fermion loops in gluon $n$-point functions have to be
resummed. We will see in the next section that this 
resummation leads to the familiar hard dense loop (HDL)
effective action \cite{Braaten:1991gm,Braaten:1992jj}.

\begin{figure}
\hbox to\hsize{\hss
\includegraphics[width=4cm]{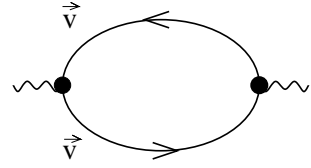}
\hss}
\caption{Hard loop contribution to the gluon
polarization function.}
\label{fig_hdl1}
\end{figure}

 The second observation is that the power counting 
for hard vertices is modified by a factor that counts
the number of fermion lines in the vertex. Using 
equ.~(\ref{dim_count}) it is easy to see that 
four-fermion operators without extra derivatives 
are leading order ($k-2-f_k=0$), but terms with 
more than four fermion fields, or extra derivatives, 
are suppressed. This result is familiar from the 
effective field theory analysis of theories with short 
range interactions \cite{Shankar:1993pf,Polchinski:1992ed}.

\section{Hard Loops}
\label{sec_hloop}

 As an example of a hard dense loop diagram we consider the 
gluon two point function. At leading order in $g$ and $1/\mu$ 
we have 
\bea 
\Pi^{ab}_{\mu\nu}(p) &=& 2g^2N_f\frac{\delta^{ab}}{2}
  \sum_{\vec{v}} v_\mu v_\nu  \nonumber \\
 & & \mbox{}\hspace{0.3cm}\cdot
 \int \frac{d^4k}{(2\pi)^4} 
 \frac{1}{(k_0-l_k)(k_0+p_0-l_{k+p})},
\eea
where $l_k=\vec{v}\cdot \vec{k}$. We note that taking the momentum
of the external gluon to be small automatically selects almost
forward scattering. We also observe that the gluon can interact 
with fermions of any Fermi velocity so that the polarization 
function involves a sum over all patches. After performing the 
$k_0$ integration we get 
\be 
\Pi^{ab}_{\mu\nu}(p) = 
  g^2N_f\delta^{ab}
  \sum_{\vec{v}} v_\mu v_\nu 
 \int \frac{d^2l_\perp}{(2\pi)^2}
 \int \frac{dl_k}{2\pi} \frac{l_p}{p_0-l_p}
  \frac{\partial n_k}{\partial l_k},
\ee
where $n_k$ is the Fermi distribution function. We note 
that the $l_k$ integration is automatically dominated
by small momenta. The integral over the transverse 
momenta combines with the sum over $\vec{v}$ as shown 
in equ.~(\ref{hard_int}). We find
\be
\label{pi_hdet2}
\Pi^{ab}_{\mu\nu}(p) = 2m^2\delta^{ab} \int\frac{d\Omega}{4\pi}
  v_\mu v_\nu \left\{ 1-\frac{p_0}{p_0-l_p}
 \right\},
\ee
with $m^2=N_fg^2\mu^2/(4\pi^2)$. This result has the correct 
dependence on $p_0,l_p$, but it is not transverse. In the 
effective theory, this can be corrected by adding a counterterm 
\cite{Hong:2000tn}
\be
{\cal L}= \frac{1}{2} m^2 \int\frac{d\Omega}{4\pi}
 (\vec{A}_\perp)^2.
\ee
The appearance of this term is related to the fact that 
the $\mu^{-1}\psi^\dagger_v D_\perp^{2}\psi_v$ vertex
gives a tadpole contribution which is not only enhanced
because of the sum over $\vec{v}$, but also by a linear 
divergence in $l_\|$. Using similar arguments one can see
that no additional counterterms are needed for $n\geq3$
point functions. Putting everything together we find
\be
\label{pi_hdet3}
\Pi_{\mu\nu}(p) = 2m^2 \int\frac{d\Omega}{4\pi}
   \left\{ \delta_{\mu 0}\delta_{\nu 0} -
   \frac{v_\mu v_\nu p_0}{p_0-l_p}
 \right\}
\ee
which agrees with the standard HDL result. The gluonic 
three-point function can be computed in the same fashion. 
We get
\bea 
\label{gam_hdet}
\Gamma^{abc}_{\mu\nu\alpha}(p,q,r) &=& igf^{abc} 2m^2 
\int \frac{d\Omega}{4\pi} v_\mu v_\alpha v_\beta 
 \nonumber \\
 & & \hspace{0.3cm}\mbox{}\cdot
  \left\{  \frac{q_0}{(q\cdot v)(p\cdot v)}
 -\frac{r_0}{(r\cdot v)(p\cdot v)}
 \right\}.
\eea
Higher order $n$-point functions can be computed in the
same way, or by exploiting Ward identities. There is 
a simple generating functional for hard dense loops
in gluon $n$-point functions which is given by 
\cite{Braaten:1991gm,Braaten:1992jj}
\be 
\label{S_hdl}
{\cal L}_{HDL} = -m^2\int\frac{d\Omega}{4\pi}
 {\rm Tr}\,G_{\mu \alpha} 
  \frac{\hat{P}^\alpha \hat{P}^\beta}{(\hat{P}\cdot D)^2} 
G^\mu_{\,\beta},
\ee
where the angular integral corresponds to an average over 
the direction of $\hat{P}_\alpha =(1,\hat{p})$. For momenta
$l<g\mu$ we have to add ${\cal L}_{HDL}$ to ${\cal L}_{HDET}$.
In order not to overcount diagrams we have to remove at the 
same time all diagrams that become disconnected if all soft 
gluon lines are deleted.

\section{Soft Loops}
\label{sec_sdl}

 As an example of a soft loop contribution in the high 
density effective theory we study the fermion self energy.
At leading order, we have
\be
\label{sigma_0}
\Sigma(p) = g^2 C_F\int\frac{d^4k}{(2\pi)^4}
  \frac{1}{p_0+k_0-l_{p+k}} v_\mu v_\nu D_{\mu\nu}(k),
\ee
where $D_{\mu\nu}(k)$ is the gluon propagator. Soft contributions
to the quark self energy are dominated by nearly forward scattering. 
Note that this loop integral does not involve a sum over patches. 
The contributions to the fermion self energy that arises from
hard loop momenta is represented by the four-fermion operators 
given in equ.~(\ref{c_flp}). 

\begin{figure}
\hbox to\hsize{\hss
\includegraphics[width=4.5cm]{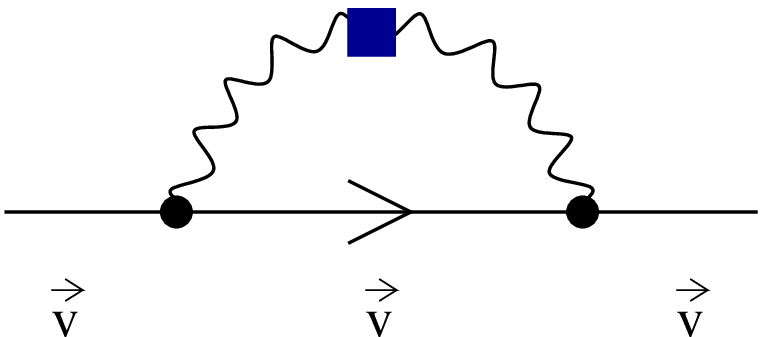}
\hss}
\vspace*{0.3cm}
\hbox to\hsize{\hss
\includegraphics[width=6cm]{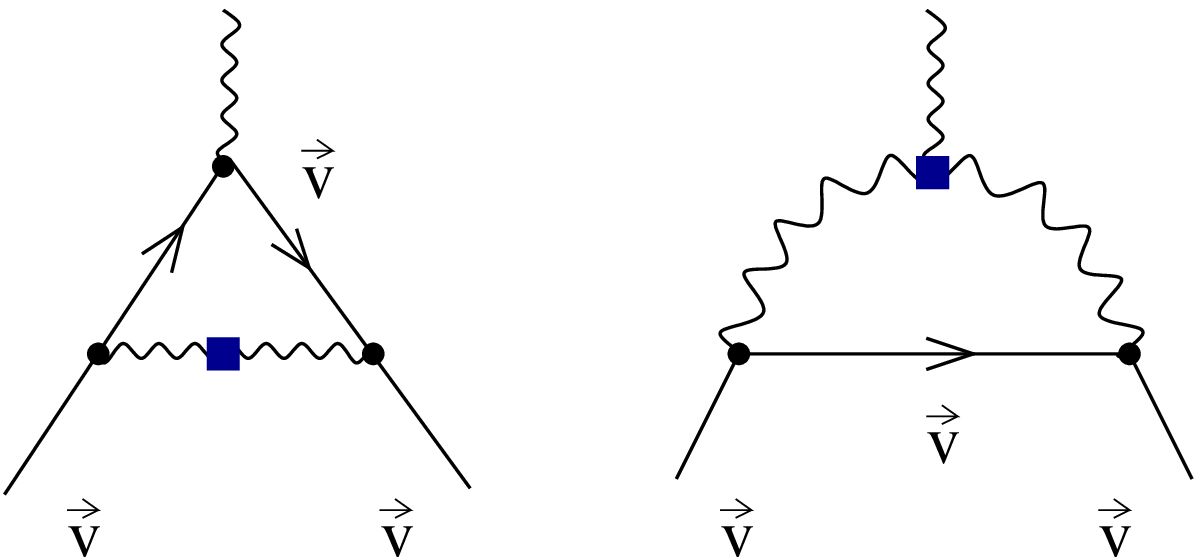}
\hss}
\caption{Leading order high density effective theory 
diagrams for the quark self energy and quark-gluon vertex
function. The solid squares indicate HDL self energy and
vertex corrections.}
\label{fig_sdl}
\end{figure}

 In the previous section we showed that for momenta $l<g\mu$
hard dense loop contributions to gluon $n$-point functions 
have to be resummed. The corresponding gluon propagator is 
given by
\be
\label{d_hdl}
D_{\mu\nu}(k) = \frac{P_{\mu\nu}^T}{k^2-\Pi_M} 
 + \frac{P_{\mu\nu}^L}{k^2-\Pi_E}
\ee
where $\Pi_M$ and $\Pi_E$ are the transverse and longitudinal 
self energies in the HDL limit. In the regime $|k_0|<|\vec{k}|
<g\mu$ the self energies can be approximated by $\Pi_E = 2m^2$ 
and  $\Pi_M = i\frac{\pi}{2}m^2k_0/|\vec{k}|$. We note that in 
this regime the transverse self energy is much smaller than the 
longitudinal one, $\Pi_M<\Pi_E$. As a consequence the dominant 
part of the fermion self energy arises from transverse gluons. 
We have 
\bea 
\label{sigma_1}
\Sigma(p) &=& g^2C_F \int \frac{dl_0}{2\pi}
 \int \frac{l^2dl}{(2\pi)^2}
 \nonumber \\
 & & \mbox{}\cdot
 \int_{-1}^1 dx \frac{1-x^2}{p_0+l_0-l_p-lx}
 \frac{1}{l_0^2-l^2+i\frac{\pi}{2}m^2 \frac{l_0}{l}},
\eea
where $l_p=\vec{v}\cdot\vec{p}-\mu$ and $l_k=\vec{v}\cdot\vec{k}
\equiv  lx$ and $C_F=(N_c^2-1)/(2N_c)$. To leading logarithmic 
accuracy we can ignore the difference between transverse and 
longitudinal cutoffs and set $\Lambda_\perp=\Lambda_{\|}=\Lambda$. We 
compute the integral by analytic continuation to euclidean space. In 
the limit $p_4\to 0$ the integral is independent of $l_p$ and given by 
\cite{Brown:1999yd,Vanderheyden:1996bw,Brown:2000eh,Manuel:2000mk,Manuel:2000nh,Boyanovsky:2000bc}
\be
\label{sigma_3}
\Sigma(p_4) \simeq \frac{g^2C_Fp_4}{4\pi^2} 
 \int  \frac{l\,dl}{l^2+\frac{\pi}{2}m^2 \frac{p_4}{l}}
 \simeq  \frac{g^2C_F}{12\pi^2}p_4 \log\left( 
  \frac{\Lambda}{p_4}\right) .
\ee
The calculation of the numerical constant inside the logarithm
requires the determination of the coefficient of the 
four-fermion operator in equ.~(\ref{c_flp}). We observe 
that the result $\Sigma\sim g^2 l$ agrees with the naive 
power counting. However, we also note that the quark 
self energy has a logarithmic divergence which spoils
the perturbative expansion for $l<\exp(-\bar{c}/g^2)$. 

 A similar logarithmic divergence appears in the quark
gluon vertex function. In order to compute this logarithm
it is essential to take into account the HDL resummed 
gluon propagator and gluon three-point function, see
Fig.~\ref{fig_sdl}. The logarithmic enhancement only 
appears in a specific kinematic configuration. We 
find \cite{Brown:2000eh}
\be 
\label{vert_sdl} 
\lim_{(p_1)_4\to (p_2)_4} \lim_{l_{p_1}\to l_{p_2}}
  \Gamma_\alpha (p_1,p_2)  = 
  \frac{g^3C_F v_\alpha}{12\pi^2}
  \log\left(\frac{\Lambda}{p_4}\right),
\ee
where $p_1,p_2$ are the momenta of the fermions. In all 
other kinematic limits the vertex correction is of order 
$g^3$, not $g^3\log(l)$.

\section{Color Superconductivity}
\label{sec_gap}

 Soft dense loop corrections to the fermion self energy 
and the quark-gluon vertex function become comparable to
the free propagator and the free vertex at the scale $E\sim 
\mu \exp(-\bar{c}/g^2)$. This implies that at this scale soft dense 
loops have to be resummed. Physically, this resummation corresponds 
to the study of non-Fermi liquid effects in dense quark matter 
\cite{Boyanovsky:2000bc,Holstein:1973,Reizer:1989,Ipp:2003cj}.
However, before non-Fermi liquid effects become important
the quark-quark interaction in the BCS channel becomes 
singular. The scale of superfluidity is $E\sim \mu\exp(-c/g)$
\cite{Son:1999uk}. 

 The resummation of the quark-quark scattering amplitude
in the BCS channel leads to the formation of a non-zero 
gap in the single particle spectrum. We can take this
effect into account in the high density effective theory 
by including a tree level gap term 
\be 
\label{gap}
{\cal L} = \Delta\, R_l^\Gamma(\vec{v}\cdot \hat\Delta)
  \psi_{-v}\sigma_2\Gamma \psi_{v} + h.c..
\ee
Here, $\Gamma$ is any of the helicity structures introduced
in Sect.~\ref{sec_hdet}, $R_l^\Gamma(x)$ is the corresponding 
angular factor and $\hat\Delta$ is a unit vector. The magnitude 
of the gap is determined variationally, by requiring the free 
energy to be stationary order by order in perturbation theory. 

 At leading order in the high density effective theory the 
variational principle for the gap $\Delta$ gives the Dyson-Schwinger 
equation 
\be 
\label{gap_1}
\Delta(p_4) =\frac{2g^2}{3} \int \frac{d^4q}{(2\pi)^4} 
 \frac{\Delta(q_4)}{q_4^2+l_q^2+\Delta(q_4)^2}
  v_\mu v_\nu D_{\mu\nu}(p-q) ,
\ee
where we have restricted ourselves to angular momentum zero and 
the color anti-symmetric $[\bar{3}]$ channel. $D_{\mu\nu}$ is 
the hard dense loop resummed gluon propagator given in 
equ.~(\ref{d_hdl}). Since the scale where soft loops 
become non-perturbative is much smaller than the scale
of superfluidity, quark self energy and vertex corrections
can be treated perturbatively. Finally, we note that 
equ.~(\ref{gap_1}) only contains collinear exchanges.
According to the arguments give in Sect.~\ref{sec_pow}
four-fermion operators are of leading order in the
HDET power counting. However, even though collinear exchanges
and four-fermion operators have the same power of $l$, 
collinear exchanges are enhanced by a logarithm of the 
small scale. As a consequence, we can treat four-fermion
operators as a perturbation. 

 We also find that to leading logarithmic accuracy
the gap equation is dominated by the IR divergence in the 
magnetic gluon propagator. This IR divergence is independent 
of the helicity and angular momentum channel. We have 
\be 
\label{gap_2}
\Delta(p_4) =\frac{g^2}{18\pi^2}
\int_0^{\Lambda_{\|}}  
\frac{\Delta(q_4)dq_4}{\sqrt{q_4^2+\Delta(q_4)^2}}
\log\left( \frac{\Lambda_\perp}{|p_4^2-q_4^2|^{1/2}}\right).
\ee
The leading logarithmic behavior is independent of the ratio 
of the cutoffs and we can set $\Lambda_{\|}=\Lambda_\perp
=\Lambda$. We introduce the dimensionless variables 
variables $x=\log(2\Lambda/(q_4+\epsilon_q))$ and $y
=\log(2\Lambda/(p_4+\epsilon_p)$ where $\epsilon_q=(q_4^2
+\Delta(q_4))^{1/2}$. In terms of dimensionless variables 
the gap equation is given by
\be 
\label{gap_3}
\Delta(y) = \frac{g^2}{18\pi^2}\int_0^{x_0} dx\, \Delta(x) K(x,y),
\ee
where $x_0=\log(2\Lambda/\Delta_0)$ and $K(x,y)$ is the kernel 
of the integral equation. At leading order we can use the 
approximation $K(x,y)=\min(x,y)$ \cite{Son:1999uk}. We can 
perform an additional rescaling $x=x_0\bar{x}$, $y=x_0\bar{y}$. 
Since the leading order kernel is homogeneous in $x,y$ we can 
write the gap equation as an eigenvalue equation 
\be 
\label{gap_4}
\Delta(\bar{y}) = x_0^2 \frac{g^2}{18\pi^2}\int_0^1 d\bar{x}\,
\Delta(\bar{x}) K(\bar{x},\bar{y}),
\ee
where the gap function is subject to the boundary conditions $\Delta
(0)=0$ and $\Delta'(1)=0$. This integral equation has the solutions 
\cite{Son:1999uk,Schafer:1999jg,Hong:2000fh,Brown:1999aq,Pisarski:2000tv} 
\be 
\label{gap_6}
\Delta_n(\bar{x}) = \Delta_{n,0} 
\sin\left( \frac{g}{3\sqrt{2}\pi}x_{0,n}\bar{x} \right), 
\hspace{0.2cm}
x_{0,n}= (2n+1)\frac{3\pi^2}{\sqrt{2}g}.
\ee
The physical solution corresponds to $n=0$ which gives the 
largest gap, $\Delta_0=2\Lambda \exp(-3\pi^2/(\sqrt{2}g))$.
Solutions with $n\neq 0$ have smaller gaps and are not 
global minima of the free energy.

\section{Higher Order Corrections to the Gap}
\label{sec_cor}

\begin{figure}
\hbox to\hsize{\hss
\includegraphics[width=2.5cm]{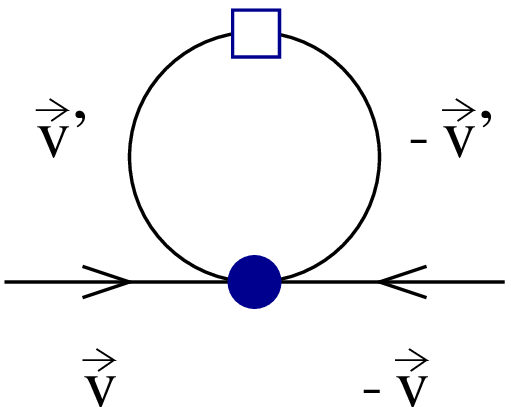}
\hss}
\vspace*{0.5cm}
\hbox to\hsize{\hss
\includegraphics[width=\hsize]{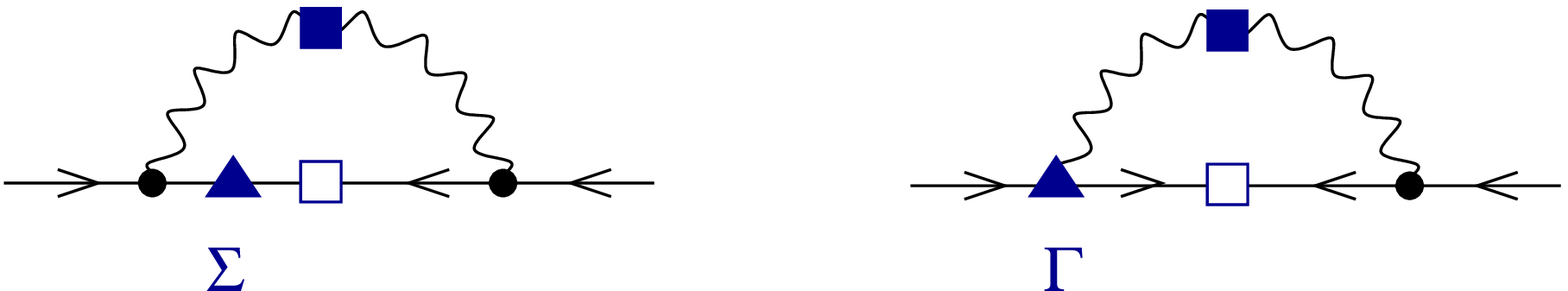}
\hss}
\caption{Higher order corrections to the gap equation in the
high density effective theory.}
\label{fig_sc2}
\end{figure}

  The high density effective field theory enables us to 
perform a systematic expansion of the kernel of the gap 
equation in powers of the small scale and the coupling 
constant. It is not so obvious, however, how to solve the 
gap equation for more complicated kernels, and how the 
perturbative expansion of the kernel is related to the 
expansion of the solution of the gap equation. 

 For this purpose it is useful to develop a perturbative method 
for solving the gap equation \cite{Brown:1999aq,Schafer:2003jn}.
We can write the kernel of the gap equation as $K(x,y)=K_0(x,y)
+\delta K(x,y)$, where $K_0(x,y)$ contains the leading IR 
divergence and $\delta K(x,y)$ is a perturbation. We expand
both the gap function $\Delta(x)$ and the eigenvalue $x_0$
order by order $\delta K$,
\bea
\Delta(\bar{x}) &=& \Delta^{(0)}(\bar{x}) + \Delta^{(1)}(\bar{x}) 
  +  \Delta^{(2)}(\bar{x}) + \ldots , \\[0.2cm]
\bar{x}_0 &=& \bar{x}_0^{(0)} + \bar{x}_0^{(1)} + \bar{x}_0^{(2)} 
  + \ldots ,
\eea
where we have defined $\bar{x}_0^2=g^2x_0^2/(18\pi^2)$. The 
expansion coefficients can be found using the fact that the 
unperturbed solutions given in equ.~(\ref{gap_6}) form an orthogonal 
set of eigenfunctions of $K_0$. The resulting expressions for 
$\bar{x}_0^{(i)}$ and $\Delta^{(i)}(\bar{x})$ are very similar
to Rayleigh-Schroedinger perturbation theory. At first order
we have
\bea
\label{dx_1}
\bar{x}_0^{(1)} &=& -\frac{1}{2}\left( \bar{x}_0^{(0)}\right)^2
\int_0^1 d\bar{x}\int_0^1d\bar{y}\, \nonumber \\
 & & \hspace{1.5cm}\cdot
\Delta_0^{(0)}(\bar{x}) 
  \delta \bar{K}(x_0\bar{x},x_0\bar{y}) \Delta_0^{(0)}(\bar{y}),
  \\
\label{dc_1}
c^{(1)}_k &=& 
 \frac{\bar{x}_0^{(0)} }{1-\left(\frac{1}{2k+1}\right)^2}   
 \int_0^1 d\bar{x} \int_0^1 d\bar{y}\, \nonumber \\
 & & \hspace{1.5cm}\cdot
 \Delta^{(0)}_0(\bar{x}) \delta \bar{K}(x_0\bar{x},x_0\bar{y}) 
 \Delta^{(0)}_k(\bar{y}),
\eea
with $\Delta^{(1)}(x)=\sum c_k^{(1)}\Delta^{(0)}_k(x)$
and $\delta\bar{K}=g/(3\sqrt{2}\pi)\delta K$.

 We can now study the role of various corrections to the 
kernel. The simplest contribution comes from collinear 
electric gluon exchanges and four-fermion operators. These
terms do not change the shape of the gap function but 
give an $O(g)$ correction to the eigenvalue $\bar{x}_0$.
This corresponds to a constant pre-exponential factor 
in the expression for the gap on the Fermi surface. An
important advantage of the effective field theory method
is that this factor is manifestly independent of the choice 
of gauge. The gauge independence of the pre-exponential
factor is related to the fact that this coefficient is 
determined by four-fermion operators in the effective 
theory, and that these operators are matched on-shell. 

  The effect of the fermion wave function renormalization is 
slightly more complicated \cite{Brown:1999aq,Wang:2001aq}. 
Using equ.~(\ref{sigma_3}) we can write
\be
\delta \bar{K}(x_0\bar{x},x_0\bar{y}) =
 -\frac{g^2}{9\pi^2} \left(\bar{x}_0\bar{x}\right)
   K_0(x_0\bar{x},x_0\bar{y}).
\ee
The corresponding correction to the eigenvalue is
\be
\label{x1_sig}
\bar{x}^{(1)}_0 = -\frac{1}{2}
 \left( \bar{x}_0^{(0)} \right)^2
 \langle 0| \delta \bar{K} | 0\rangle 
 =\frac{4+\pi^2}{8}\frac{g}{3\sqrt{2}\pi}, 
\ee
where $\langle 0| \delta \bar{K} | 0\rangle $ denotes the 
matrix element of the kernel between unperturbed gap 
functions, see equ.~(\ref{dx_1}). At this order in $g$, 
there is no contribution from the quark-gluon vertex 
correction. 

 Note that the quark self energy correction makes an 
$O(g)$ correction to the kernel, even though it is an
$O(g^2)$ correction to the kernel. This is related to 
the logarithmic divergence in the self energy. The 
perturbative expansion of $\bar{x}_0$ is of the form
\be
\bar{x}_0 \sim g\log(\Delta) 
 = O(g^0) + O(g\log(g)) + O(g) + \ldots .
\ee
Brown et al.~argued that equ.~(\ref{x1_sig}) completes the 
$O(g)$ term. The result for the spin zero gap in the 2SC 
phase at this order is \cite{Brown:1999yd,Wang:2001aq,Schafer:2003jn} 
\be
\Delta = 512\pi^4 \mu g^{-5}
 e^{-\frac{4+\pi^2}{8}}e^{-\frac{3\pi^2}{\sqrt{2}g}}.
\ee
In other spin or flavor channels the relevant four fermion
operators are different and the pre-exponential factor is 
modified \cite{Brown:1999yd,Schafer:2000tw,Schmitt:2002sc}. 

\section{Very Low Energies} 
\label{sec_gb}

  For momenta below the gap the dynamics is determined
by Goldstone modes. In the CFL phase the effective 
lagrangian of the form \cite{Casalbuoni:1999wu}
\bea
\label{l_cheft}
{\cal L}_{eff} &=& \frac{f_\pi^2}{4} {\rm Tr}\left[
 \partial_0\Sigma\partial_0\Sigma^\dagger - v_\pi^2
 \partial_i\Sigma\partial_i\Sigma^\dagger \right] 
 +\Big\{ B {\rm Tr}(M\Sigma^\dagger) \nonumber  \\ 
 & & \hspace*{-1cm}\mbox{} 
     + A_1\left[{\rm Tr}(M\Sigma^\dagger)\right]^2 
     + A_2{\rm Tr}\left[(M\Sigma^\dagger)^2\right]
     \nonumber \\
 & & \hspace*{-1cm}\mbox{} 
     + A_3{\rm Tr}(M\Sigma^\dagger){\rm Tr} (M^\dagger\Sigma)
         + h.c. \Big\}+\ldots . 
\eea
Here $\Sigma=\exp(i\phi^a\lambda^a/f_\pi)$ is the chiral field,
$f_\pi$ is the pion decay constant and $M$ is a complex mass
matrix. The chiral field and the mass matrix transform as
$\Sigma\to L\Sigma R^\dagger$ and  $M\to LMR^\dagger$ under 
chiral transformations $(L,R)\in SU(3)_L\times SU(3)_R$. We 
have suppressed the singlet fields associated with the breaking 
of the exact $U(1)_V$ and approximate $U(1)_A$ symmetries.
The coefficients $f_\pi,B,A_i$ can be determined by matching 
the effective chiral lagrangian to the high density effective 
theory \cite{Son:1999cm,Schafer:2002ty,Schafer:2001za}

 The chiral expansion has the structure
\bea
{\cal L} \sim f_\pi^2\Delta^2
\left( \frac{\vec\partial}{\Delta} \right)^{k}
\left( \frac{\partial_0+MM^\dagger/p_F}{\Delta} \right)^{l}
\left( \frac{MM}{p_F^2} \right)^{m}
(\Sigma)^{n}
(\Sigma^\dagger)^{o}
\eea
Loop graphs are suppressed by powers of $p/(4\pi f_\pi)$.
Since the pion decay constant scale as $f_\pi\sim p_F$ loops 
are parametrically small as compared to higher order contact 
terms. The quark mass expansion is somewhat subtle because of 
the appearance of two scales, $m^2/p_F^2$ and $m^2/(p_F\Delta)$.
This problem is discussed in more detail in \cite{Bedaque:2001je}.

\section{Conclusions} 
\label{sec_con}

 In this contribution we discussed effective
field theories in QCD at high baryon density. We focused,
in particular, on the problem of power counting in the
high density effective theory. We showed that the power
counting is complicated by ``hard dense loops'', i.e.
loop diagrams that involve the large scale $\mu^2$. We
proposed a modified power counting that takes these
effects into account. The modified $l$ counting implies
that hard dense loops in gluon $n$-point functions have
to be resummed below the scale $g\mu$, and that four
fermion operators are leading order in the HDET power
counting. There are a number of important questions that 
remain to be addressed. An example is the renormalization 
of operators in the high density effective field theory.

 Acknowledgments: We would like to thank I.~Stewart for useful 
discussions. This work was supported in part by US DOE grant 
DE-FG-88ER40388.

\end{document}